# ON RATE OF UNIVERSE EXPANSION

**Lennur Ya. Arifov**[1]

Results of measurements of dependence z on photometric distances $d_L$ to supernovae SNe Ia carried out on the Hubble Space Telescope revealed failure of standard cosmological models for modern epoch of Universe evolution. In order to adjust the theoretical curve with experimental one in the frames of standard models it was necessary to change the Einstein equations. Selection of special value for cosmological constant $\Lambda$ allowed to restore the harmony but in the frames of $\Lambda + \mathrm{CDM}$ standard model. Now the measurements of function $d_L(z)$ is taken as experimental proof of existence of dark energy with the density several times larger comparing with the average density of ordinary matter. According to the $\Lambda$+CDM model negative pressure of dark energy caused accelerated expansion of the Universe. This induced the author to think more carefully about the bases of standard cosmological models and propose another method for description of modern epoch of Universe in the frames of classic Einstein's equations.

## 1. INTRODUCTION

Use of the supernovae SNe Ia as a standard candle (Branch & Tammann 1992) of out-of-galaxy objects allowed to increase measuring accuracy of dependence of objects photometric distance $d_L$ on shift of their spectral lines $z$. As a result of measurements of function $d_L(z)$ in the interval $z[0, 0.9]$ and theoretical interpretation of obtained data Riess et al. (1998), Garnavich et al. (1998), Perlmutter et al. (1999) came to fundamental for cosmology conclusion about accelerated expansion of the Universe. It means that exotic forms of gravitational interaction sources with energetic characteristics $r + 3p \le 0$ (here $r -$ mass-energy density, $p -$ pressure) dominate over other for sure known forms with equation of state satisfying the condition $r + 3p > 0$. And what is more, if standard models are adequate to modern state of the Universe then the results of measurements of $d_L(z)$ for SNe Ia are to be interpreted as experimental proof of existence of hypothetical before form of dark energy (Weinberg 1989; Peebles 1999; Coldwell & Steinhardt 1998; Wang et al. 2000), which is connected with cosmological constant $\Lambda$ (equation of state $r_\Lambda + p_\Lambda = 0$, $r_\Lambda = const.$) or other hypothetical fields (e.g. quintessence with equation of state $-r \le p \le -\frac{1}{3} r$ ).

For this moment[2] acknowledged cosmological model is $\Lambda + \mathrm{CDM}$ model: $k = 0$ (flat space), $\Omega_\Lambda \approx 0.6 \div 0.7$, $\Omega_d \approx 0.3$, $\Omega_b \approx 0.03$, $\Omega = \sum \Omega_n = 1$, $\mathrm{H}_0 \sim 65\ \mathrm{km \cdot s^{-1} \cdot Mpc^{-1}}$. Here are used standard cosmological notations for Hubble constant and relations $\Omega_n$ of mass density $r_n(t_0)$ of corresponding form of the source to $r_c = \frac{3}{k}\mathrm{H}_0{}^2$; $r_d -$ mass density of dark matter, $r_b -$ mass density of baryonic matter, $k -$ Einstein's gravitational constant, $t_0 -$ modern epoch.

[1] Taurida National V.I. Vernadsky University, avenue V.I. Vernadsky 4, 95007 Simferopol, Ukraine. E-mail arifov @ tnu.crimea.ua

[2] According to expression Wang et al. (2000) in cosmological society the cosmology concordance is accepted.

We do not call in question results of measurements of function $d_L(z)$ for SNe Ia. But we take note that negative value of deceleration parameter $q_0$ (corresponds to accelerated expansion of the Universe) is not a straight consequence of measuring results, but follows from their theoretical interpretation. Real result of the measurements is the detection of nonlinear dependence of registered energy current density of SNe Ia on shift parameter of their spectral lines $z$. The character of this nonlinear dependence is compatible with the theory of function $d_L(z)$ in the standard models only if equations of state of dominating component from all possible sources of gravitational field satisfies the condition $r + 3p < 0$. In that way there is a dilemma: either bases of standard models are wrong or in addition to the for sure known forms of gravitational field sources the Universe is filled with other forms with exotic equations of state and mass-energy density several times of typical matter density. These forms define the character of Universe evolution. In the last case negative pressure acts as gravitational repulsive forces, and results of experiments are interpreted as accelerated expansion of the Universe. In $\Lambda + \mathrm{CDM}$ standard model this dilemma is solved in favour of traditional homogenous isotropic models introduced in primary resulting articles (Friedmann 1922, 1924), dedicated to the construction of cosmological models on the basis of Einstein equations.

Taking into account the fundamental role of value $q_0 < 0$ for cosmology and physics as a whole, we think it is quite suitable to consider critically those of basic cosmological propositions which are the base for interpretation of experimental data and theory of function $d_L(z)$. We investigate this problem in the following two sections. In sections 4 – 6 we state alternative model of evolutional stage A of the Universe and derivation of theoretical formula for $d_L(z)$. This formula adjusts with experimental curve without any additional parameters or exotic energy forms (see section 7). The Universe decelerates while expanding. Deceleration parameter $q_0$ equals to $\Omega_r$ ($r_r$ – cosmic microwave background energy density), and space warp of Universe is minus quantity because $k = -1$.

## 2. BASIC PROPOSITIONS OF STANDARD MODELS

Let's concentrate on the following basic propositions of standard cosmological models:

1. Space-time is homogeneous and isotropic. Its geometry is defined by quadratic form (Robertson 1933)

$$ds^2 = -dt^2 + a^2\left[dc^2 + A^2\left(dq^2 + \sin^2 q\, dj^2\right)\right], \tag{1a}$$

$$a = a(t), \quad A = \frac{1}{2\sqrt{-k}}\left(e^{y} - e^{-y}\right), \quad y = \sqrt{-k}\,c\,, \quad k = -1, 0, +1\,, \qquad \text{(1b, c, d, e)}$$

$c\,, q\,, j$ – radial (dimensionless) and angular coordinates of points on the sphere.

2. Distribution of gravitational field sources is continues, homogeneous and isotropic. It is described by stress-energy tensor of multicomponent perfect fluid filling the space with mass energy density $r(t)$ and pressure $p(t)$.

3. In the reference frame co-moving by the perfect fluid Einstein equations are reduced to the following form:

$$3\left(\frac{\dot{a}^2}{a^2}+\frac{k}{a^2}\right)=k\,r\ ,\quad 2\frac{\ddot{a}}{a}+\frac{\dot{a}^2}{a^2}+\frac{k}{a^2}=-k\,p\ . \tag{2a, b}$$

Here and further dot over the symbol denotes the derivative of the corresponding quantity with respect to time $t$.

4. From all known components of perfect fluid main contribution to the mass energy density comes from baryonic and dark matter. The most investigated baryonic matter is concentrated mainly in the stars of the galaxies and partly in interstellar gas-dust medium. Its mass density is evaluated by $r_b(t_0)\approx 2\times10^{-31}\,\mathrm{g\cdot cm^{-3}}$. Dark matter cannot be observed directly. But according to many independent indirect measurements estimation (Olive 2002) of dark mass density gives the value $r_d(t_0)\approx 2\times10^{-30}\,\mathrm{g\cdot cm^{-3}}$. Dark mass is also concentrated in galaxies and galactic clusters. That is why it consists of elements with non zero rest mass.

Energy density of cosmic microwave background (CMB) radiation is measured much better $r_r(t_0)=4.47\times10^{-34}\ \mathrm{g\cdot cm^{-3}}$. Spatial energy distribution of CMB radiation is homogeneous and isotropic, relative value of inhomogeneity $\sim10^{-5}$ (Sazhin 2004). Its temperature at present epoch $\approx 2.7\ \mathrm{K}$, temperature of baryonic and dark matter is much lower. That is why on wide enough interval of values $z[0\,,100]$ we can neglect the pressure of baryonic and dark matter and take $p_b=p_d=0$. Pressure of CMB radiation is defined by its density $p_r=\frac{1}{3}r_r$.

5. Neutrino is the other one of for sure known particles of permanent component cosmic perfect fluid. If rest mass of neutrino is equal to zero, then $r_n\sim r_r$. However rest mass of neutrino must be not equal to zero. Experiments recently carried out on solar and atmospheric neutrinos oscillations gave limitations on mass of the neutrino (Fukuda et al. 1998) not only upside but also below: $\left\{1.5<\left|m_2{}^2-m_1{}^2\right|<5\right\}\times10^{-3}\,\mathrm{eV}^2$, here $m_1$ and $m_2$ − mass of neutrinos of different species. If these data are true ( Koshiba 2004; Wark 2003; Bettini 2001), then neutrinos of cosmic origin are part of dark matter and contribute into the value of $r_d(t_0)$.

6. In traditional standard cosmological models only two components of perfect fluid $r_b$ and $r_r$ are taken into account. In CDM standard models dark matter component $r_d$ is added to these components. But in $\Lambda+\mathrm{CDM}$ models cosmic perfect fluid consists of 4 and more components. Additional components are hypothetical and characterized by exotic equation of state $r+3p\le0$.

7. For 4-componential perfect fluid

$$r=r_b+r_d+r_r+r_\Lambda\ ,\quad p_b=p_d=0\ ,\quad p_\Lambda=-r_\Lambda\ , \tag{3}$$

under the assumption of constant entropy of the Universe per unit particle function $\mathrm{d_L}(z)$ is defined by accurate formulas

$$\mathrm{d_L}=(1+z)\,a_0A(c_0)\ , \tag{4}$$

$$c_0 = a_0^{-1} H_0^{-1} \int_1^{1+z} \left[ \Omega_\Lambda + (1-\Omega) x^2 + \Omega_m x^3 + \Omega_r x^4 \right]^{-1/2} dx \;, \tag{5}$$

where

$$\Omega_m = \Omega_b + \Omega_d \;, \qquad a_0 = a(t_0) \,, \qquad \Omega = \Omega_\Lambda + \Omega_m + \Omega_r = 1 + k\, a_0^{-2} H_0^{-2} . \tag{6a, b, c, d}$$

If put $k = 0$, then formulas (4)-(6) are reduced to the following form:

$$\mathrm{d_L} = H_0^{-1} (1+z) \int_1^{1+z} \left[ \Omega_\Lambda + \Omega_m x^3 + \Omega_r x^4 \right]^{-1/2} dx \quad , \qquad \Omega_\Lambda + \Omega_m + \Omega_r = 1 . \tag{7a, b}$$

In CDM standard models $\Omega_\Lambda = 0$, and formulas (4)-(6) can be reduced to:

$$\begin{aligned} \mathrm{d_L} = 2 H_0^{-1} &\left[ 4\Omega_r - (\Omega_m + 2\Omega_r)^2 \right]^{-1} \\ &\times \left\{ (2 - \Omega_m - 2\Omega_r) \sqrt{1 + (\Omega_m + 2\Omega_r) z + \Omega_r z^2} - 2(1-\Omega_r) + \Omega_m (1-z) \right\} , \end{aligned} \tag{8}$$

independently of values $k = -1,\ 0, +1$.

Comparison of formulas (7) and (8) with data of measurements $\mathrm{d_L}(\mathrm{z})$ on Hubble Space Telescope (Riess et al. 1998; Garnavich et al. 1998; Perlmutter et al. 1999) showed that formula (8) gives understated values with augmentation of $z$. Conclusion is that the results of measurements testify in favour of $r_\Lambda \sim (0.6 \div 0.7)\, r_c$, and correspondingly about the accelerated expansion of the Universe.

## 3. CRITICAL ANALISIS OF BASIC PROPOSITIONS OF STANDARD MODELS

States 2 and 3 are under doubt. State 2 replaces real matter distribution of the Universe by homogeneous multicomponent thermodynamic system. Such a replacement in cosmology is called matter “homogenization” (Weinberg 1972). State 3 for real matter distribution in Einstein equations is replaced by equations (2) for homogenized matter. These replacements need to be proved and can have a right to exist if they do not contradict observational data.

Einstein equations

$$\mathbf{G}(\mathbf{g}) = -k\ \mathbf{T}(\mathbf{g}), \tag{9}$$

are local and establish accurate equality between geometrical characteristic of space-time and energetic characteristic of gravitational field sources in every point of space at every time. Here $\mathbf{G}$ – Einstein tensor, $\mathbf{g}$ – metric space-time tensor, $\mathbf{T}$ – stress-energy tensor. Einstein and stress-energy tensors are defined on given metric tensor.

If we turn to the evolution of the Universe, then in the past we can conventionally distinguish several significantly different stages of evolution (Fig. 1). Probability of the Universe to be in stages C and B is quite high and follows from extrapolation into past of to known observational data: expansion of the Universe and non zero homogeneous isotropic CMB radiation in the Universe at present time. Stage C is characterized by high temperature of gravitational field sources and high value of energy density, matter is the state of electrically neutral plasma. Due to the expansion of the Universe structure of the plasma changes with time from elemental particles of different types on early stages of evolution C to electrons, neutrinos, hydrogen, helium, deuterium atomic nucleus and other chemical elements on late stages. Matter on that stage comprises the elements of future dark matter. As plasma and electromagnetic radiation were in thermodynamic equilibrium plasma should have homogeneous and isotropic mass and energy distribution with small deviation $\sim 10^{-5}$ from homogeneity. That is why it is possible to modulate right hand side of equations (9) by homogeneous equilibrium system in stage C. Form (1) in this case is the one which reveals the symmetry – homogeneous and isotropy - of the left hand side of equations (9). Therefore, standard cosmological models adequately describe stage C in the evolution of the Universe.

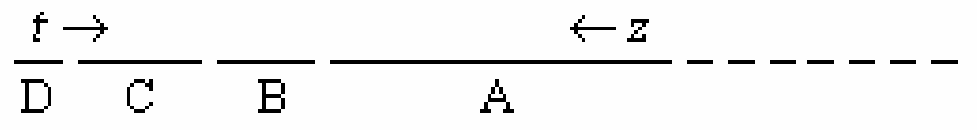

D - stage of initial expansion of the Universe from singular state
C - stage of Friedmann expansion of homogeneous exuilibrium thermodynamic system
B - transitional stage of matter structurization and appearence of multiphased system
A - expansion stage of formed multiphased system. Dotted line denotes evolution of stage A of the Universe into the future

**Fig. 1**

Stage A radically differs from stage C. Distribution of gravitational filed sources is quite inhomogeneous in stage A. Here we have complicated system where defined enough and for sure possible to distinguish two thermodynamic phases and with under special conditions third phase[1]. Radiation phase stipulated mainly by CMB radiation occupies the most part of the Universe space. Relative value of local deviations from inhomogeneity is $\sim 10^{-5}$.

Stellar phase is an aggregate of stars and similar to them gravitationally tied local objects of the Universe. It is multiply connected and has quite precise bounds separating it from radiation phase. Average value of stars number density $< n >_U$ in the Universe $\sim 3\times 10^{-9}\,\text{pc}^{-3}$; average value of mass density $< r_b >_U \approx 2\times 10^{-31}\,\text{g cm}^{-3}$; volume of stellar phase to volume of radiation phase is $\sim 10^{-31}$. Mass density inside an object of stellar phase changes in $10^{10}$ times and even more, but stars number density changes from zero in intergalactic void (their volume is $\sim 10^{23}\,\text{pc}^3$) to $10^7\,\text{pc}^{-3}$ in the center of the galaxies. Surface temperature of stellar phase much more exceeds the temperature of radiation phase.

Galactic phase is an aggregate of galaxies, groups and galactic clusters. It comprises as a part of a system stellar phase, dark matter of the Universe and interstellar matter inside galaxies. Galaxies number density in the Universe changes from zero in the intergalactic void to $\sim 300\,\text{Mpc}^{-3}$ in the center of Abell's galactic clusters, and their average value is $\sim 0.01\,\text{Mpc}^{-3}$. Average mass density of galactic phase in the Universe is $< r_m >_U \sim 2\times 10^{-30}\,\text{g cm}^{-3}$, while average mass density in the galaxy $< r_m >_g \sim \left(10^{-22} \div 10^{-23}\right)\text{g cm}^{-3}$. That is why volume of galactic phase is $\sim 10^{-7} \div 10^{-8}$ of the volume of the Universe.

Galaxies and galactic clusters are gravitationally tied local objects and have precise bounds. But the bounds of galactic phase do not coincide with the bounds of radiation phase. The bounds of radiation phase are surfaces of stellar phase and surfaces of ionized gas clouds inside galaxies. So separation of galactic phase is conventional and is due to convenience of analysis of space matter distribution. Exactly galactic phase constitute large-scale structure of the Universe

[1]Note that introduced here description of stages A and C of Universe evolution does not pretend to detail and accuracy of numbers. Matter distribution and its structure are simplified in the extent necessary and sufficient for drawing well-grounded conclusions about correctness of application of Einstein equations in standard cosmological models in the form (2a, b).

(Oort 1983). Signs of homogeneity reveals in this structure in the scale $> 200$ Mpc. Lower limit of homogeneity scale according to different estimations (Tarakanov 2005) belongs to the wave band from 200 Mpc to $10^3$ Mpc. Notice that right limit of this band can already be compared with the quantity $L_U$ of cosmological horizon $\left(L_U \sim 10^4 \text{ Mpc}\right)$. For wave lengths less 200 Mpc there are no any signs of inhomogeneity of galactic phase. For homogenization of real matter distribution according to state 2 of standard models existence of scale $l$ satisfying the condition $l \ll L_U$ is demanded, and degree of homogeneity of matter has to be not less then $10^{-5}$ (according to the degree of homogeneity of radiation phase).

Assume that as a result of data's treatment scale $l$ is found. After averaging of stress-energy tensor on scale $l$ it is necessary to make double substitution in the right hand side of equations (9)

$$\mathbf{T}(\mathbf{g}) \Rightarrow <\mathbf{T}(\mathbf{g})>_l \Rightarrow \mathbf{T}_{i.l.}(<\mathbf{g}>_l). \tag{10}$$

Here $\mathbf{T}_{i.l.}$ – stress-energy tensor of perfect fluid, and $<\mathbf{g}>_l$ corresponds to the metric tensor of form (1). Then equations (2a, b) used in standard models coincide with equations

$$\mathbf{G}(<\mathbf{g}>_l) = -k\ \mathbf{T}_{i.l.}(<\mathbf{g}>_l) \quad , \tag{11}$$

but do not coincide with the averaged Einstein equations

$$<\mathbf{G}(\mathbf{g})> = -k <\mathbf{T}(\mathbf{g})>. \tag{12}$$

Therefore we can define several problems without solution of which application of standard models for description of evolutionary stages A and B of the Universe is not correct.

1. Simple definition of value $<\text{X}>$ for tensor quantities in Riemann geometry in the absence of space symmetry. (Definition of $<\text{X}>$ for scalar functions is out of doubt.)

2. Substantiation of equality

$$<\mathbf{T}(\mathbf{g})>_l = \mathbf{T}_{i.l.}(<\mathbf{g}>_l) \tag{13}$$

for real matter distribution in stage A of the Universe.

3. Proof of equality

$$<\mathbf{G}(\mathbf{g})> = \mathbf{G}(<\mathbf{g}>) \tag{14}$$

for arbitrary metric tensor.

It seems first problem has not a solution. Summation (integration) of tensor quantities excepting scalar functions defined on some region of Riemann space has not single meaning. But definition of $<\text{X}>$ is based on summation. Such problems already were discussed in general relativity while trying to formulate integral laws of energy, impulse and angular momentum conservation. As it is know satisfactory solution to this problem was not found. Second and third problems are connected with the first one. Proof of equation (14) because of high degree of nonlinearity of function $\mathbf{G}(\mathbf{g})$ is not simple task even if first problem is somehow solved. To establish a connection between Einstein equations in the form of (9) or (12) and equations (2a, b) of standard models without proving equality (14) is not possible.

Stage B of the evolution of the Universe is transitional. After recombination of negative and positive electric charges of plasma in stage C radiation phase separates from neutral matter, fluctuation of matter density increases and it structures. Description of dynamics of the matter in this stage demands kinetic methods and correspondingly transforms Einstein equations. Argumentation for applying standard models in stage B is even less with compared to stage A.

## 4. EVOLUTIONARY STAGE OF THE UNIVERSE

In the evolutionary stage A of the Universe we can distinguish quite definitely two forms of gravitationally tied objects comprising stellar and galactic phases with precise bounds. Denote part of space inside these bounds as $D_m$. The rest part of space denote as $D_r$. Defining characteristics of regions $D_m$ and $D_r$ are inequality $\mathbf{T}_m\left(D_m\right) \neq 0$ for $D_m$ and equality $\mathbf{T}_m\left(D_r\right) = 0$ for $D_r$. Here $\mathbf{T}_m = \mathbf{T}_b + \mathbf{T}_d$, $\mathbf{T}_b$ – stress-energy tensor of baryonic matter, and $\mathbf{T}_d$ – stress-energy tensor of dark matter. In $D_r$ stress-energy tensor a $\mathbf{T}_r$ of radiation phase is not equal to zero. It is equal to zero in some part of region $D_m$. Ratio of volumes $V\left(D_m\right) \cdot V^{-1}\left(D_r\right) \sim 10^{-7} \div 10^{-8}$, and ratio of mass densities $r_m \cdot r_r^{-1} \sim 4\times10^3$. That is why for definition of the dynamics of the Universe in stage A it is natural to come to classical statement of boundary value problem for Einstein equations:

$$\mathbf{G} = -k\left(\mathbf{T}_m + \mathbf{T}_r\right), \quad P \in D_m ; \tag{15a}$$

$$\mathbf{G} = -k\mathbf{T}_r , \quad P \in D_r ; \tag{15b}$$

$$\mathbf{G} = -k\left(\mathbf{T}_m + \mathbf{T}_r\right), \quad P \in D_m \bigcap D_r ; \tag{15c}$$

here $P$ – space-time event.

Equations (15a) and (15b) distinguish essentially from physical point of view. Mass density of matter in stellar phase exceeds value of mass density in galactic phase much more, and the last exceeds mass density of radiation phase. That is why equations (15a) define the role of gravitational field in the intrinsic structure of stars, in motion of stars and another matter in the galaxies and motion of galaxies in galactic clusters relatively centre of inertia. While homogeneous radiation phase fills the whole space of the Universe. Let us affirm that qualitative and mostly quantitative characteristics of dynamics of the evolution of the Universe in stage A are defined by equations (15b). Centers of inertia of galactic clusters are at rest in the referenced frame co-moving by the radiation phase.

## 5. EINSTEIN EQUATIONS FOR RADIATION PHASE

Energy of radiation phase is continuously distributed in space and is not strictly homogeneous. There are two reasons of its inhomogeneity. First reason – evolutionary. Fluctuations of energy density of plasma and electromagnetic field were in stage C conserved by radiation phase after its separation from matter in stage B and transformed during expansion of the Universe. The other reason is the gravitational interaction of radiation phase with matter of galactic phase. Inhomogeneous of radiation phase is not compatible with matter inhomogeneity in stellar and galactic phases in stage A because of equations (15c), sewing together gravitational field on the bound of space regions $D_m$ and $D_r$. Estimation of homogeneity of this nature may be dimensionless ratio of mass to linear sizes of objects of galactic phase. For most of the galaxies this ratio is within the limits of $10^{-5} \div 10^{-7}$ .

Stress- energy tensor of radiation phase has a form

$$T_{rb}^{a} = \frac{1}{3} r_r \left( d_b^a + 4u^a u_b \right).$$

Here $u^a$ – time-like proper vector $\mathbf{T}_r$, directed in the future and satisfied the condition $u^a u_a = -1,$ Greek indices run from 0 to 3.

In co-moving reference system

$$u^a u_b = -\frac{g_{0b}}{g_{00}} \cdot d_0^a \ , \qquad (16a)$$

that is why

$$T_{rb}^{a} = \frac{1}{3} r_r \left( d_b^a - 4\frac{g_{0b}}{g_{00}} d_0^a \right), \qquad (16b)$$

where $g_{ab}$ – metric tensor.

Assume

$$r_r = < r_r > + e \ , \qquad (17a)$$

$< r_r >$ – homogeneous component of energy density of radiation phase, function of dimensionless time coordinate $h$ , (17b)

$e(h, M)$ – function $h$ and point $M$ on the hypersurface $h = const.$ (17c)

Additive representation (17) of right hand side of Einstein equations (15b) is corresponded to the multiplicative representation of quadratic form of space-time

$$ds^2 = b^2 \left\{ -e^{2\Phi} dh^2 + 2e^{\Phi} g_k dx^k dh + h_{kn} dx^k dx^n \right\} \qquad (18)$$

for left hand side (15b).

Here $b = b(h)$, $\Phi$ , $g_k$ , $h_{kn}$ – dimensionless functions $h$ and dimensionless space coordinates $x^i$; Latin indices run from 1 to 3.

Denote metric tensor of the form in brackets in the right hand side of equation (18) by $g_{ab}$ so that

$$g_{00} = -e^{2\Phi}, \quad g_{0k} = e^{\Phi} g_k , \quad g_{kn} = h_{kn} . \qquad (19a, b, c)$$

Correspondingly $G_{ab}$ и $\Gamma_{ab}^{l}$ – Einstein tensor and Christoffel symbols, defined on metric tensor $g_{ab}$ . Metric tensor of the form (18) is now defined as $b^2 \cdot g_{ab}$ . Stress-energy tensor in representation of (16) is invariant with regard to substitution $g_{ab} \to b^2 \cdot g_{ab}$ . Substituting (16), (17) and metric tensor of form (18) into equations (15b) reduce them (see Appendix) to:

$$G_{ab} - L_{ab} = -\frac{1}{3} k e b^2 \left( g_{ab} - 4\frac{g_{0a} g_{0b}}{g_{00}} \right), \qquad (20)$$

$$b = \frac{a}{\sqrt{-k}} e^{-\frac{x}{2}} \cdot \left(1 - e^x\right), \tag{21}$$

$$< r_r > \; = \; 12 k^2 k^{-1} a^{-2} e^{2x} \cdot \left(1 - e^x\right)^{-4} . \tag{22}$$

Here

$$L_{ab} = 2\left(k + 2l^2\right) d_a^0 d_b^0 + 2l\,\Gamma_{ab}^0 + 4\left(l^2 + k\right) \frac{g_{0a} g_{0b}}{g_{00}} \\ - g_{ab}\left[k + l^2 + \left(l^2 + 2k\right) g^{00} + 2l\, g^{mn}\,\Gamma_{mn}^0\right]; \tag{23}$$

$$l = \sqrt{-k}\left(1 + e^x\right)\left(1 - e^x\right)^{-1}, \qquad k = -1, 0, +1 \quad ; \tag{24a, b}$$

$$x = 2\sqrt{-k}\left(h_s - h\right) \;, \quad h_s = const. \quad a = const. \tag{25a, b, c}$$

Equations (20) are accurate, and define metric tensor $g_{ab}$ in region $D_r$ as function of inhomogeneities $e\left(h, M\right)$ of radiation phase energy density distribution. Solution of equations (20) continuously depends on values $e\left(h, M\right)$ on the initial hypersurface $h = 0$ and boundary $D_m \cap D_r$. According to the theorem (see Appendix), metric tensor $g_{ab}$ satisfying equation (20), also satisfies the condition

$$\lim_{e \to 0} g_{0a} = -d_a^0 \,, \qquad \lim_{e \to 0} \frac{\partial g_{kn}}{\partial h} = 0 \,, \tag{26a, b}$$

And quadratic form (18) in the limit $e = 0$ is equivalent to the form

$$ds^2 = b^2\left\{-dh^2 + dc^2 + A^2\left(dq^2 + \sin^2 q\, dj^2\right)\right\}, \tag{27}$$

$$A = \frac{1}{2\sqrt{-k}}\left(e^y - e^{-y}\right), \qquad y = \sqrt{-k}\; c \;, \tag{28a, b}$$

accurate to coordinates transformation which do not disturb co-moving character of reference frame.

Constants $a$, $h_s$ and a value of $k$ are defined by experimental values of Hubble constant $H_0$ and of energy density $< r_r >_0$ at the modern epoch $h = 0$.

## 6. PHOTOMETRIC DISTANCE FUNCTION IN THE STAGE A OF THE UNIVERSE

The Hubble's effect theory and derivation of photometric distance function $d_L$ is based on the assumption that the observer and the object emitting light registered by the observer are at rest in the reference frame co-moving by the radiation phase. Both the observer and object belong to the region $D_m$, and light distributes in $D_r$ region. Information about spectral lines shifts of cosmological origin enclosed exactly in properties of region $D_r$ of the Universe. But it is overburden by Doppler effect (stipulated by speeds of the object and observer relatively centers of inertia gravitationally tied subsystems $D_m$, where they locate) and gravitational shift of spectral lines (stipulated by local gravitational fields in $D_m$), that is why will be estimated with large mistake. Anyway relative errors $\mathrm{H}_0$ in modern estimations are $10^{-1}$. Magnitude of this ratio $\left[e \cdot < r_r >^{-1}\right]_{h=0} \sim 10^{-5}$. That is why geometrical and physical properties of radiation phase in stage A of the Universe with necessary accuracy are revealed by formulas (21), (22) and (27).

Use the formula for definition of the frequency of electromagnetic wave

$$w = w_0 \cdot \left|u_a k^a\right|, \qquad u_a = -\frac{1}{b} d_a^0 \; ; \tag{29a, b}$$

where $\mathbf{k}$ – is tangent vector of isotropic geodesics, $k^a k_a = 0$, $w_0$ – scale factor.
In space- time (27) we have

$$k^a = \frac{1}{b^2} \cdot k_0{}^a \;, \qquad \left(k_0{}^a\right)' = 0 \,, \qquad g_{ab} k_0{}^a k_0{}^b = 0. \tag{30a, b, c}$$

Here and further $X' = \partial_h X$ .
Locate space coordinate origin on the object's world line $\Gamma_1$ (see Fig.2), we get for isotropic geodesics

$$c = h + c_0 \;, \quad q = q_0, \quad j = j_0. \tag{31a, b, c}$$

From (29) and (30) follows

$$\frac{w_1}{w_2} = \frac{b_0}{b(h)} = 1 + z \; . \tag{32a, b}$$

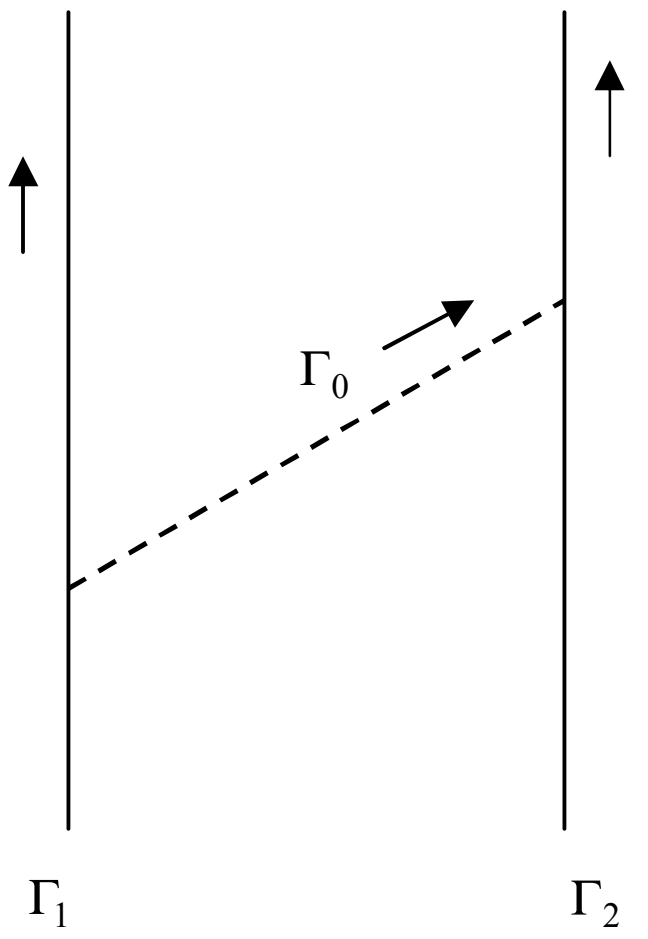

$\Gamma_1$ – world line of emitting object ( $c = 0$ )
$\Gamma_2$ – world line of observer $\left(c = c_0 \,, j = j_0 \,, q = q_0\right)$
$\Gamma_0$ – isotropic geodesic.

**Fig. 2**

Here $w_1$ – frequency of light emitted by object on $\Gamma_1$; $w_2$ – frequency of light registered by the observer on $\Gamma_2$; $b_0 = b_{|h=0}$.
When $z \ll 1$, which corresponds to $c_0 \ll 1$, $|h| \ll 1$, $l(\Gamma_1, \Gamma_2) \approx b_0 c_0$, Hubble effect follows from formula (32b)

$$z \approx \mathrm{H}_0 l \,, \qquad \mathrm{H}_0 = \frac{b'}{b^2}_{|h=0} \; . \tag{33a, b}$$

Substitute equation (32b) into eq. (A8d, e) and taking into account eq. (A3), get

$$k = \left(\Omega_r - 1\right) b_0{}^2 \mathrm{H}_0{}^2 \, . \tag{34}$$

Comparing eq. (34) with the results of estimations of $< r_r >_0$ and $\mathrm{H}_0$

$$< r_r >_0 = 4.40\times10^{-34}\ \mathrm{g\,cm^{-3}}, \qquad \left(50 < \mathrm{H}_0 < 100\right)\mathrm{km\,s^{-1}\,Mpc^{-1}}, \tag{35a, b}$$

$$2.34\times10^{-5} < \Omega_r < 9.36\times10^{-5}\ , \tag{35c}$$

find that

$$k = -1, \qquad b_0 = \mathrm{H}_0^{-1}\left(1-\Omega_r\right)^{-1/2}\ . \tag{36a, b}$$

Now constants $a$ and $h_s$ are expressed through $\Omega_r$ и $\mathrm{H}_0$

$$h_s = arch\frac{1}{\sqrt{\Omega_r}}\ , \qquad a = \frac{1}{2}\frac{\sqrt{\Omega_r}}{\mathrm{H}_0\left(1-\Omega_r\right)}\ ,$$

and formulas (21), (22), (24 a) and (28 a) reduce to the following form:

$$b = \frac{1}{\mathrm{H}_0\left(1-\Omega_r\right)}\left\{shh + \sqrt{1-\Omega_r}\ chh\right\}, \tag{37a}$$

$$< r_r > = \frac{3}{k}\mathrm{H}_0^{\,2}\,\Omega_r\left(1-\Omega_r\right)^2\left[shh + \sqrt{1-\Omega_r}\,chh\right]^{-4}, \tag{37b}$$

$$l = \left(chh + \sqrt{1-\Omega_r}\ shh\right)\left(shh + \sqrt{1-\Omega_r}\ chh\right)^{-1}, \tag{37c}$$

$$A = sh\,c\ . \tag{37d}$$

As we do not have any other distinguished time $t$ in stage A put $t = 0$ for modern epoch. Then

$$t = \frac{1}{\mathrm{H}_0\left(1-\Omega_r\right)}\left(chh + \sqrt{1-\Omega_r}\ shh - 1\right)\ , \tag{38a}$$

$$t_s = -\frac{1}{\mathrm{H}_0}\,\frac{1-\sqrt{\Omega_r}}{1-\Omega_r}\ . \tag{38b}$$

From equations (37a) and (38) follows

$$\dot{b} > 0, \quad \ddot{b} < 0 \quad \text{on the interval } t_s < t < \infty\ ,$$

$$\lim_{t\to\infty}\dot{b} = 1\ , \qquad \lim_{t\to\infty} b^3\ddot{b} = -\,\Omega_r\mathrm{H}_0^{-2}\left(1-\Omega_r\right)^{-2}\ .$$

Consequently expansion of the Universe in stage A takes place with deceleration. Parameter of deceleration is equal to $\Omega_r$.

If $\mathrm{L}$ – the luminosity of the object on the world line $\Gamma_1$ (see Fig. .2), then photometric distance $\mathrm{d_L}$ from observer on line $\Gamma_2$ to the object is defined by the equality

$$d_L = (1+z) b_0 A(c_0), \qquad c_0 = -h . \tag{39a, b}$$

Solve equality (32b) relatively function $h(z)$ and substitute into eq. (39) and (38a), get

$$d_L = \frac{1}{H_0 \Omega_r}\left[\sqrt{1+z(2+z)\Omega_r} \quad -1\right], \tag{40}$$

$$t = \frac{1}{H_0 (1-\Omega_r)} (1+z)^{-1}\left[\sqrt{1+z(z+2)\Omega_r} \quad -1-z\right], \tag{41}$$

$$t_{|z=0} = 0, \qquad \lim_{z\to\infty} t = t_s .$$

Upper limit of changes of $z$ we can define from the condition $< r_m >_U \approx < r_m >_g$. From equality (32 b) follows that changes of z on interval $[0 \to 100]$ corresponds to space volume decrease and increase $< r_m >_U$ in $10^6$ times. Therefore when $z \sim 100$ ratio $< r_m >_U \cdot < r_m >_g^{-1} \sim 1$. This value z we can assume as conventional boundary between evolutionary stages A and B of the Universe. Duration $\Delta t_A$ of stage A is defined by formula (41) subject to equations (35c).

$$\Delta t_A \sim 0.98 \frac{1}{H_0} \quad . \tag{42}$$

When $z > 100$, this model is not applicable.

## 7. COMPARISON OF PHOTOMETRIC DISTANCE FUNCTION WITH EXPERIMENTAL DATA

For comparison of theoretical function $d_L(z)$, defined by formula (40), with experimental data numerical results of measurements of dependence of $z$ on energy flow density coming from supernovae SNe Ia – standard candle - are demanded. As we do not have them we will use formulas (7) as an experimental function $d_L(z)$. Formula (7a) is accurate theoretical formula in $\Lambda + CDM$ model and adjusts with the measuring results (Riess et al. 1998; Garnavich et al. 1998; Perlmutter et al. 1999) in the interval $z[0, 0.9]$. Therefore denoting right hand side as $d_L{}^\Lambda$, rewrite (7a) as:

$$d_L{}^\Lambda = \frac{f^\Lambda(z)}{H_0{}^\Lambda}, \qquad f^\Lambda = (1+z)\int_1^{1+z}\left[1+\Omega_m\left(x^3-1\right)+\Omega_r\left(x^4-1\right)\right]^{-\frac{1}{2}} dx , \tag{43a, b}$$

$$H_0{}^\Lambda = (65 \pm 7)\ \mathrm{km\,s^{-1}Mpc^{-1}}, \qquad \Omega_m = 0.33, \qquad \Omega_r = 5.63 \cdot 10^{-5}. \tag{43c, d, e}$$

If introduce designation

$$b \; = \; \mathrm{H_0}^{\Lambda} \cdot \left(\mathrm{H_0}^{\mathrm{A}}\right)^{-1} \, , \qquad (44)$$

then formula (40) can be rewritten in the following form:

$$\mathrm{d_L}^{\mathrm{A}} = f^{\mathrm{A}} \cdot \left(\mathrm{H_0}^{\Lambda}\right)^{-1} \, , \qquad f^{\mathrm{A}} = \frac{1}{\Omega_r b}\left[\sqrt{1+\mathrm{z}(2+\mathrm{z})\Omega_r b^2} \;\; -1\right]. \qquad (45\text{a, b,})$$

Choosing parameter $b$ we minimize root-mean-square deviation of function $f^{\mathrm{A}}$ and $f^{\Lambda}$ in the interval $\mathrm{z}[0\,,\,1]$, and get

$$b = 1.0294, \qquad \int_0^1 \left(f^{\mathrm{A}} - f^{\Lambda}\right)^2 dx = 0.848 \cdot 10^{-4} \, , \qquad d \; = \; \frac{0.92}{\mathrm{H_0}^{\Lambda}} \cdot 10^{-2} \, , \qquad (46\text{a, b, c})$$

where $d$ – root-mean-square deviations of functions $\mathrm{d_L}^{\mathrm{A}}$ и $\mathrm{d_L}^{\Lambda}$.
Interval $\mathrm{z}[0\,,\,1]$ is chosen here not accidentally. Values of $\mathrm{z}$ of supernovae SNe Ia, discovered in the experiments on Hubble Space Telescope and used (Riess et al. 1998; Garnavich et el. 1998; Perlmutter et al. 1999) for investigation of function $\mathrm{d_L}(\mathrm{z})$, do not exceed 0.9.

Values of functions $f^{\mathrm{A}}(\mathrm{z})$ for $b = 1.0294$ and $f^{\Lambda}(\mathrm{z})$ in the interval $\mathrm{z}[0\,,\,1]$ are shown in the Table, and in the interval $\mathrm{z}[0\,,\,3]$ – in the Fig. 3. From table data follows that in the interval $z[0\,,\,1]$ they coincide in three points when $z \approx 0.15$ and $z \approx 0.81$. In the interval $z[0\,,\,0.15]$ function $f^{\mathrm{A}}$ exceeds $f^{\Lambda}$, then in the interval $z[0.15\,,\,0.81]$ it goes under $f^{\Lambda}$, and in the interval $z[0.81\,,\,1]$ again is larger than $f^{\Lambda}$. Break between them continues to increase while $z$ grows in the interval $z > 1$. Relative value of deviation of functions $\mathrm{d_L}^{\mathrm{A}}$ and $\mathrm{d_L}^{\Lambda}$ coincide, according to formulas (43a) and (45a), with relative deviation value of functions $f^{\mathrm{A}}$ and $f^{\Lambda}$, therefore from table data (see also Fig. 4) get

$$-0.0183 < \Delta < 0.0286 \quad \text{in the interval} \quad \mathrm{z}[0\,,\,1]\,, \qquad \Delta \; = \; \frac{\mathrm{d_L}^{\mathrm{A}} - \mathrm{d_L}^{\Lambda}}{\mathrm{d_L}^{\Lambda}}. \qquad (47\text{a, b})$$

Substituting eq. (46a) into eq. (44), find the value of Hubble constant $\mathrm{H_0}^{\mathrm{A}}$ corresponding to “data processing” of measuring results of photometric distances SNe Ia (Riess et al. 1998; Garnavich et al. 1998; Perlmutter et al. 1999) following the formula (40).

$$\mathrm{H_0}^{\mathrm{A}} = 63.14 \;\mathrm{km\,s^{-1}Mpc^{-1}}\,, \qquad \frac{\mathrm{H_0}^{\Lambda} - \mathrm{H_0}^{\mathrm{A}}}{\mathrm{H_0}^{\Lambda}} = 2.86 \cdot 10^{-2} \, . \qquad (48\text{a, b})$$

Thus formula (40) conforms to formula (7a) for $d_L(z)$ in the interval $\mathrm{z}[0\,,\,1]$ in $\Lambda + \mathrm{CDM}$ model, and consequently – with experimental data also. Formulas (40) and (7a) essentially differ when $\mathrm{z} > 1$.

## Table

| $z$ | $f^{A}$ | $f^{\Lambda}$ | $\Delta$ |
|---|---|---|---|
| .00 | 0 | 0 | .28560e-1 |
| .02 | .2079e-1 | .20299e-1 | .23617e-1 |
| .04 | .4200e-1 | .41185e-1 | .19405e-1 |
| .06 | .6362e-1 | .62646e-1 | .15310e-1 |
| .08 | .8564e-1 | .84669e-1 | .11338e-1 |
| .10 | .10808 | .10724 | .77720e-2 |
| .12 | .13093 | .13035 | .44298e-2 |
| .14 | .15421 | .15397 | .15563e-2 |
| .16 | .17788 | .17809 | -.11806e-2 |
| .18 | .20197 | .20272 | -.37134e-2 |
| .20 | .22647 | .22783 | -.60052e-2 |
| .22 | .25138 | .25342 | -.81152e-2 |
| .24 | .27670 | .27946 | -.99747e-2 |
| .26 | .30244 | .30594 | -.11573e-1 |
| .28 | .32858 | .33286 | -.13026e-1 |
| .30 | .35514 | .36020 | -.14248e-1 |
| .32 | .38211 | .38796 | -.15310e-1 |
| .34 | .40949 | .41612 | -.16191e-1 |
| .36 | .43729 | .44468 | -.16900e-1 |
| .38 | .46550 | .47362 | -.17444e-1 |
| .40 | .49411 | .50292 | -.17830e-1 |
| .42 | .52315 | .53260 | -.18064e-1 |
| .44 | .55256 | .56262 | -.18206e-1 |
| .46 | .58242 | .59298 | -.18131e-1 |
| .48 | .61268 | .62369 | -.17970e-1 |
| .50 | .64336 | .65472 | -.17657e-1 |
| .52 | .67445 | .68607 | -.17229e-1 |
| .54 | .70594 | .71773 | -.16701e-1 |
| .56 | .73786 | .74969 | -.16033e-1 |
| .58 | .77018 | .78194 | -.15269e-1 |
| .60 | .80291 | .81448 | -.14410e-1 |
| .62 | .83606 | .84731 | -.13456e-1 |
| .64 | .86962 | .88040 | -.12396e-1 |
| .66 | .90359 | .91378 | -.11277e-1 |
| .68 | .93797 | .94740 | -.10054e-1 |
| .70 | .97275 | .98127 | -.87587e-2 |
| .72 | 1.00795 | 1.0154 | -.73413e-2 |
| .74 | 1.04358 | 1.0498 | -.59410e-2 |
| .76 | 1.07961 | 1.0844 | -.44461e-2 |
| .78 | 1.11603 | 1.1192 | -.28674e-2 |
| .80 | 1.15289 | 1.1543 | -.12143e-2 |
| .82 | 1.19016 | 1.1896 | .50412e-3 |
| .84 | 1.22783 | 1.2251 | .21991e-2 |
| .86 | 1.26591 | 1.2608 | .40288e-2 |
| .88 | 1.30440 | 1.2967 | .59031e-2 |
| .90 | 1.34331 | 1.3328 | .78166e-2 |
| .92 | 1.38264 | 1.3691 | .97642e-2 |
| .94 | 1.42237 | 1.4056 | .11811e-1 |
| .96 | 1.46251 | 1.4423 | .13812e-1 |
| .98 | 1.50306 | 1.4792 | .15900e-1 |
| 1.00 | 1.54402 | 1.5162 | .18005e-1 |

In the first column are shown values of spectral lines shift - $z$ , in the second and third columns - values of functions $f^{A}(z)$ and $f^{\Lambda}(z)$ correspondingly, in the forth column – values of relative deviation $\Delta(z)$ of photometric distance functions $d_L{}^{A}$ and $d_L{}^{\Lambda}$ .

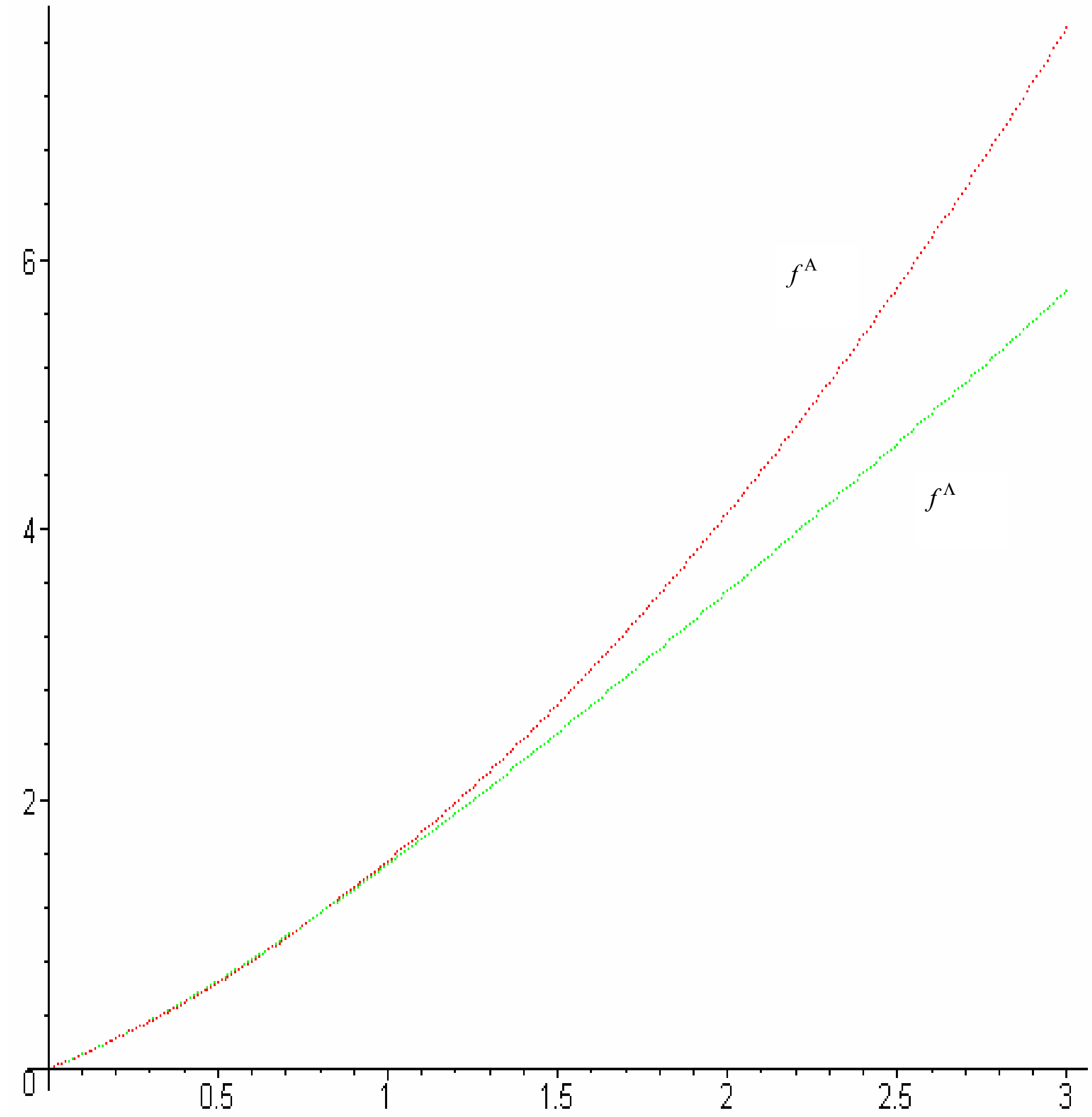

On the axis of abscissas values of $z$ are shown. Red line – function $f^{A}$, adjusted with function $f^{\Lambda}$ – green line - by the method of least squares in the interval $z\,[0\,,1]$

**Fig. 3**

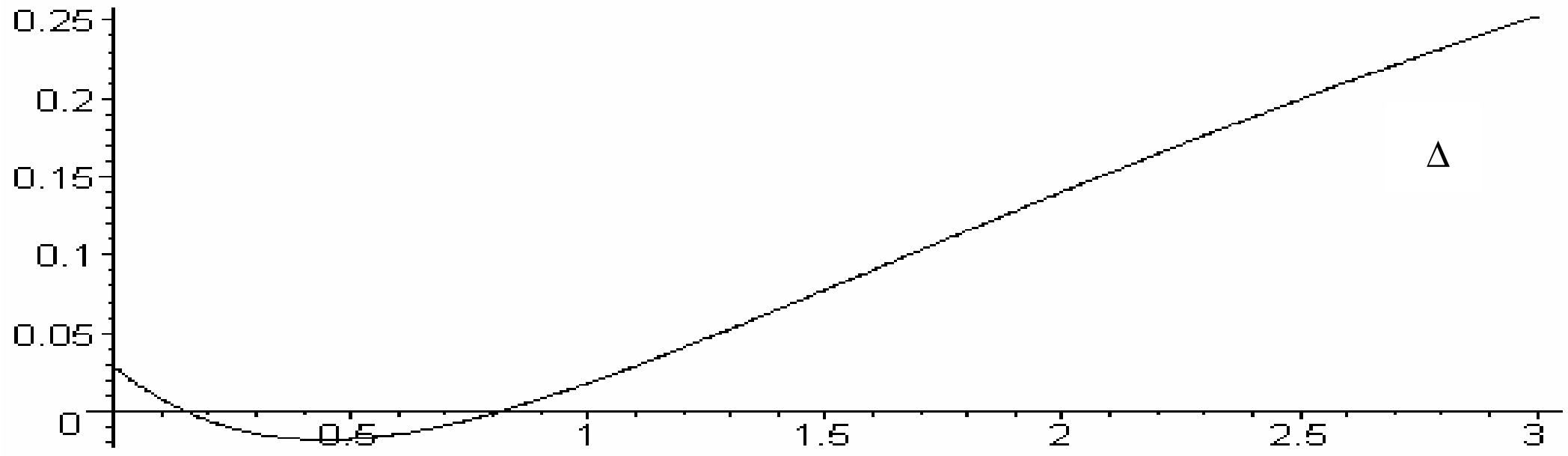

On the axis of abscissas values of $z$ are shown. $\Delta$ – function of relative deviation value of photometric distances $d_L{}^{A}$ and $d_L{}^{\Lambda}$. Values of $|\Delta|$ in the interval $z[0\,,1]$ are the smaller of 0.03. In the interval $z > 1$ function $\Delta(z)$ increases till 1.

**Fig. 4**

## 8. CONCLUSIONS

1. Baryonic and dark matter distribution in the modern epoch of the evolution of the Universe is not homogeneous. Therefore use of Einstein equations in the form traditionally used in standard cosmological models - since publication of Friedman's papers (1922, 1924) – can not be acknowledged to be correct. Replacement of real matter and radiation distribution by "homogenized" matter – perfect fluid – in the right hand side of Einstein equations assumes corresponding averaging of left hand side. Here is hidden the reason of impossibility to describe theoretically nonlinear dependence of function $d_L(z)$ received in measurements on Hubble Space Telescope in the frames of CDM ($\Lambda = 0$) standard models. Introducing special term interpreted as vacuum energy in classical Einstein equations and selecting value of additional parameter $\Lambda$ it was possible to adjust with the theoretical curve in the interval $z[0\,,0.9]$. When the interval increases (measurements of photometric distances of more distant supernovae SNe Ia) we can expect new problems to appear.

2. Standard models adequately describe evolutionary stage C of the Universe when real matter and radiation distribution under high temperature with high degree of precision are homogeneous equilibrium thermodynamic system. Alternative to standard models in stage A including modern epoch is either transition from Einstein equations to averaged ones[1], or statement of boundary value problem for Einstein equation. In this work an account of variant of second alternative is given.

3. Alternative model is based on the following observational data:

3.1. Presence of the CMB radiation in the Universe – radiation phase. Energy density of radiation phase $r_r = < r_r > + e$ with high degree of precision $e \cdot < r_r >^{-1} \sim 10^{-5}$ is homogeneous, but its kinematical functions are isotropic and homogeneous.

3.2. Matter (baryonic and dark) constitutes multiply connected large-scaled structure elements of which are local gravitationally tied objects (galactic clusters). Volume of space $V_g$ occupied by this structure constitutes $\sim 10^{-7} \div 10^{-8}$ of radiation phase volume.

3.3. Centers of inertia of large-scale structure members are at rest in the reference frame co-moving by the radiation phase.[2]

4. Within the scope of this model was received a function $d_L(z)$ precision of which is defined by degree of homogeneous accuracy of radiation phase energy distribution and exceeds by several order of magnitude the accuracy of modern measurements of dependence of $d_L$ on $z$. It satisfactorily adjusts with the results of measurements (Riess et al. 1998; Garnavich et al. 1998; Perlmutter et al. 1999) of photometric distances of supernovae SNe Ia.

5. Expansion of the Universe in the evolutionary stage A including modern epoch occurs with deceleration. Parameter of deceleration is equal to $\Omega_r$. Duration of stage A $\sim 0.98 \times H_0^{-1}$. Space warp is minus quantity ($k = -1$).

6. In contrast to $\Lambda + \mathrm{CDM}$ model in the model suggested here there are no any additional excepting Hubble constant $H_0$ adjusting parameters.

---

[1] In the history of physics the analogues is such a transition is averaging of Maxwell-Lorentz equations in the medium, thereof they transform into Maxwell equations.

[2] Strictly speaking, this statement is not direct consequence of observational data. It follows from the fact of Universe expansion in the modern epoch, evolutionary connection of stages A and C and formation of large-scale structure after separation of radiation phase from matter under temperature drop.

**APPENDIX**

Assume that $h\left(=x^0\right)$ and $x^i\left(=x^i\right)-$ are dimensionless time and space coordinates. Space-time quadratic form without commonness constrains can be written as

$$ds^2 = b^2 \cdot g_{ab}\, dx^a\, dx^b\ , \tag{A1}$$

where $b=b(h)$, $g_{00}=-e^{2\Phi}$, $g_{0k}=e^{\Phi}g_k$, $g_{kn}=h_{kn}$,

$\Phi$, $g_k$, $h_{kn}$ – coordinate functions $h$ and $x^k$.

Introduce denotes $\Gamma^m_{ab}$, $R^a_{\cdot bmn}$, $R_{ab}$, $G_{ab}$, $R$ respectively for Christoffel symbols, Riemann – Christoffel, Ricci, Einstein tensors and scalar curvature defined on metric tensor $g_{ab}$. Then Einstein tensor defined on metric tensor $b^2 \cdot g_{ab}$, is equal to

$$G_{ab}+2\left(l'-l^2\right)d^0_a d^0_b - 2l\,\Gamma^0_{ab} + g_{ab}\left[2l\,g^{mn}\,\Gamma^0_{mn} - \left(2l'+l^2\right)g_{00}\right], \tag{A2}$$

where

$$l=\frac{b'}{b}\ , \tag{A3}$$

Here and further stroke denotes derivative of corresponding value on time coordinate $h$.

In the co-moving reference system stress-energy tensor of radiation phase defined on metric $b^2\cdot\mathbf{g}$, according to (16b) is

$$T_{(r)ab}=\frac{1}{3}r_r\,b^2\left(g_{ab}-4\frac{g_{0a}\,g_{0b}}{g_{00}}\right). \tag{A4}$$

Replacing left hand side of equation (15b) by expression (A2) and taking into account of formula (A4) in the right hand side, get the following equations

$$\begin{aligned} &G_{ab}+2\left(l'-l^2\right)d^0_a d^0_b - 2l\,\Gamma^0_{ab} \\ &\quad + g_{ab}\left[2l\,g^{mn}\,\Gamma^0_{mn} - \left(2l'+l^2\right)g^{00}\right] = -\frac{1}{3}k\,r_r b^2\left(g_{ab}-4\frac{g_{0a}\,g_{0b}}{g_{00}}\right). \end{aligned} \tag{A5}$$

If $h'_{kn}=0$, then $h_{kn}$ represents metric tensor $\mathbf{h}$ of space $R^3$ in coordinates $x^k$. Introduce the following notations $g^{\,i}_{kn}$, $r^i_{\cdot knj}$, $r_{kn}$ and $r$ correspondingly for Christoffel symbols, Riemann – Christoffel, Ricci, tensors and scalar curvature defined on $\mathbf{h}$ in $R^3$. If $h'_{kn}\neq 0$, then $\mathbf{h}$ and any function defined on it in $R^3$, depend on $h$ as on parameter.

Let now define a function$e\left(h,M\right)$ in space-time with quadratic form (A1)

$$\partial_i e = \partial_i r_r\,, \qquad r_r = < r_r > + \; e\,, \qquad M \in R^3\,. \tag{A6}$$

Function $< r_r >$ является однородной составляющей $r_r$. Note that relatione $e \cdot < r_r >^{-1}$ is not necessary small.

**Lemma.** Conditions

$$g_k = 0, \qquad h'_{kn} = 0, \tag{A7a, b}$$

are necessary and sufficient in order to satisfy the equalities

$$\Phi = 0, \qquad e = 0\;; \tag{A8a, b}$$

$$r_{ijkn} = k\left(h_{ik}h_{nj} - h_{in}h_{jk}\right), \qquad k = -1, 0, +1\;; \tag{A8c}$$

$$l^2 + k = \frac{1}{3}k\,b^2 < r_r >; \tag{A8d}$$

$$l' = -\frac{1}{3}k\,b^2 < r_r >; \tag{A8e}$$

$$w^a = 0\,, \qquad w_{ab} = 0\,, \qquad s_{ab} = 0\,, \qquad J = 3\frac{b'}{b^2}\,. \tag{A8f, g, h, i}$$

Here were used the notations for kinematical characteristics (Synge 1960) of radiation phase: $w^a$ – absolute acceleration vector, $w_{ab}$ –rotation tensor, $s_{ab}$ –tensor of shear deformations velocities, $J$ –relative velocity of volume change.

***Sufficiency.*** From conditions (A7a, b) follow the equations

$$\Gamma^0_{00} = \Phi'\,, \quad \Gamma^i_{00} = e^{2\Phi}h^{ik}\partial_k\Phi\,, \quad \Gamma^0_{0k} = \partial_k\Phi\,, \quad \Gamma^0_{kn} = \Gamma^i_{0k} = 0\,, \quad \Gamma^i_{kn} = g^{\,i}_{kn}\;; \tag{A9a-f}$$

$$R^0_{\cdot k0i} = -\partial^2_{ik}\Phi + g^{\,n}_{ik}\,\partial_n\Phi - \partial_i\partial_k\Phi\,, \tag{A10a}$$

$$R^k_{\cdot 00i} = -e^{2\Phi}h^{kn}\left(\partial^2_{in}\Phi - g^{\,j}_{in}\,\partial_j\Phi + \partial_i\Phi\,\partial_n\Phi\right), \tag{A10b}$$

$$R^i_{\cdot knj} = r^i_{\cdot knj}\,, \quad \text{other components are equal to zero;} \tag{A10c, d}$$

$$R_{00} = -e^{2\Phi}h^{in}\left(\partial^2_{in}\Phi - g^{\,k}_{in}\,\partial_k\Phi + \partial_i\Phi\,\partial_n\Phi\right), \qquad R_{0k} = 0\,, \tag{A11a, b}$$

$$R_{ik} = r_{ik} + \partial^2_{ik}\Phi - g^{\,n}_{ik}\,\partial_n\Phi + \partial_i\Phi\,\partial_k\Phi\,. \tag{A11c}$$

Substituting equalities (A7a), (A9b) and (A11b) into the equations $(0k)$ of system (A5), get

$$l \cdot \partial_k \Phi = 0 , \quad \Rightarrow \quad \partial_k \Phi = 0 , \tag{A12}$$

As $l \neq 0$. Equalities (A9)-(A11) reduce now to the following:

$$\Gamma^0_{00} = \Phi' , \qquad \Gamma^i_{kn} = g^{\,i}_{kn} , \quad \text{the others are equal to zero;} \tag{A13a, b, c}$$

$$R^i_{\cdot knj} = r^i_{\cdot knj} , \quad \text{the others are equal to zero;} \tag{A14a, b}$$

$$R_{kn} = r_{kn} , \qquad R_{00} = R_{0k} = 0 , \qquad R = r . \tag{A15a-d}$$

Substituting them into other equations $(00)$ and $(kn)$ of system (A5), get

$$r = 2\left(3l^2 e^{-2\Phi} - k\, r_r b^2\right), \tag{A16a}$$

$$r_{kn} = 2e^{-2\Phi}\left( l\,\Phi' - l' + l^2 - \frac{2}{3} k\, r_r b^2 e^{2\Phi} \right) h_{kn} . \tag{A16b}$$

Make use of formula

$$r_{ijkn} = h_{in} r_{kj} + h_{kj} r_{in} - h_{ik} r_{nj} - h_{nj} r_{ik} - \frac{1}{2} r\left(h_{in} h_{kj} - h_{ik} h_{nj}\right), \tag{A17}$$

correct in every $R^3$. Then equalities (A16a, b) can be rewritten in the form

$$r_{ijkn} = K\left(h_{ik} h_{jn} - h_{in} h_{jk}\right), \qquad K = const., \tag{A18a}$$

$$l^2 = \left(\frac{1}{3} k\, r_r b^2 - K\right) e^{2\Phi} ; \tag{A18b}$$

$$l' - l\,\Phi' = -\frac{1}{3} k\, r_r b^2 e^{2\Phi} . \tag{A18c}$$

As since $b = b(h)$ by definition and $l$ and $\Phi$ do not depend on space coordinates according to equalities (A3) and (A12) correspondingly, then from equation (A18c) follows equality (A8b).

Using coordinate transformations

$$h \to \tilde{h} = \sqrt{|K|}\int e^{\Phi} dh , \qquad x^k \to \tilde{x}^k = \sqrt{|K|}\, x^k , \quad \text{when } K \neq 0 , \tag{A19a, b}$$

$$h \to \tilde{h} = \int e^{\Phi} dh , \quad \text{when } K = 0 , \tag{A19c}$$

we can simultaneously calibrate function $\Phi(h)$ and constant curvature $K$ of space $R^3$ to the values

$$\Phi = 0\,, \qquad K \to k = -1, 0, +1\,. \tag{A20a, b}$$

This calibration is incorrect because transformations (A19) do not change the character of reference frame. After coordinate transformations it stays co-moving as it is assumed in Einstein equations (A5).

Equalities (A18) in calibration (A20) coincide with eq. (A8a, c, d, e). Use now the definition of kinematical functions of CMB radiation. In the co-moving reference frame they reduce to the following equalities:

$$w_0 = 0\,, \qquad w_k = \left( g_k{}' + \frac{b'}{b} g_k \right) e^{-\Phi}\,; \tag{A21a, b}$$

$$w_{0a} = 0\,, \qquad w_{kn} = \frac{1}{2} b\left(\partial_n g_k - \partial_k g_n + g_n w_k - g_k w_n\right); \tag{A22a, b}$$

$$s_{0a} = 0\,, \qquad s_{kn} = \frac{b}{2} e^{-\Phi}\left[ \left(h_{kn} + g_k g_n\right)' + \frac{1}{3}\left(h_{kn} + g_k g_n\right)\left( \frac{s'}{s} - \frac{h'}{h} \right) \right]; \tag{A23a b}$$

$$J = \frac{1}{b}\left( 3\frac{b'}{b} - \frac{1}{2}\frac{s'}{s} + \frac{1}{2}\frac{h'}{h} \right) e^{-\Phi}\,. \tag{A24}$$

Here

$$s = \left(1 + g_k g^k\right), \qquad g^k = h^{kn} g_n\,, \qquad h = \left|h_{kn}\right|. \tag{A25a, b, c}$$

Taking into account lemma conditions (A7) and equality (A20a) we can easily derive equalities (A8f, g, i) from eq. (A21)-(A24).

□

***Necessity.*** Equality (A7b) follows from eq. (A8c). Equating now left hand sides of equations (A21b) and (A23b) to zero, according to eq. (A8f) and (A8h), and taking into account eq. (A8a), get

$$g_k' + \frac{b'}{b} g_k = 0 \tag{A26}$$

and

$$\frac{s'}{s}\left(h_{kn} + g_k g_n\right) + 3\left(g_k g_n\right)' = 0\,. \tag{A27}$$

Combining these equations it is easy to make sure that $s = 1$ and $g_k = 0$.

□

Note that space $R^3$, according (A8c), is homogeneous and isotropic. Homogeneous component of radiation phase density energy $< r_r >$ and its three isotropic (A8f, g, h) and one homogeneous (A8i) kinematical functions correspond to it. That is why physical meaning of lemma is in the fact that conditions (A7) are sufficient and necessary for derivation of simple solution of equations (A5), which adjusts with isotropic and homogeneous component of its right hand side.

General solution of equations (A8d, e) defines functions $b(h)$ and $< r_r >$:

$$b = \frac{a}{\sqrt{-k}} e^{-\frac{x}{2}}\left(1-e^{x}\right), \qquad < r_r > = \frac{12k^2}{ka^2}\frac{e^{2x}}{\left(1-e^{x}\right)^4}, \tag{A28a, b}$$

where

$$x = 2\sqrt{-k}\left(h_s - h\right), \qquad h_s = const., \qquad a = const.$$

When $h \to h_s$ function $< r_r >$ increases infinitely, and $b(h)$ – decreases to zero

$$< r_r > \to \frac{3}{4ka^2\left(h_s - h\right)^4}, \qquad b \to -2a\left(h_s - h\right).$$

That is why singular stage in the evolution of the Universe formally corresponds to values $h \sim h_s$. But it is worth keep in our mind that this model is applicable only to the radiation phase in stage A when $h \gg h_s$.

Function $l\left(z\right)$ is now defined by equality

$$l = \sqrt{-k}\left(1+e^{x}\right)\left(1-e^{x}\right)^{-1}. \tag{A29}$$

Substituting equations (A8d, e) into eq. (A5), rewrite them in the following form:

$$G_{ab} - L_{ab} = -\frac{1}{3}keb^2\left(g_{ab} - 4\frac{g_{0a}\,g_{0b}}{g_{00}}\right); \tag{A30}$$

$$\begin{aligned} L_{ab} = {} & 2\left(k+2l^2\right)d_a^0 d_b^0 + 2l\,\Gamma_{ab}^0 \\ & + 4\left(k+l^2\right)\frac{g_{0a}\,g_{0b}}{g_{00}} - g_{ab}\left[k + l^2 + \left(2k+l^2\right)g^{00} + 2l\,g^{mn}\,\Gamma_{mn}^0\right]. \end{aligned} \tag{A31}$$

Equations (A30) are accurate in region $D_r$, and metric tensor $g_{ab}$, satisfying to the equations (A30), continuously depend on initial values of $e_0\left(M\right) = e\left(0, M\right)$ on the hypersurface $h = 0$, $M \in R^3$, and on boundary conditions of $e\left(h, M\right)$ on $D_r \cap D_m$.

Now we can formulate the following theorem.

**Theorem.** If radiation phase energy distribution $r_r$ has homogeneous component $< r_r >$ and its motion in the limit $e = 0$ is homogeneous and isotropic, i.e.

$$\lim_{e \to 0} w^a = 0, \qquad \lim_{e \to 0} w_{ab} = 0, \qquad \lim_{e \to 0} s_{ab} = 0, \qquad \lim_{e \to 0} \partial_k J = 0, \qquad \text{(A32a, b, c, d)}$$

then solution of equations (A30) satisfies the conditions

$$\lim_{e \to 0} g_{0a} = -d_a^0 , \qquad \lim_{e \to 0} g'_{kn} = 0 , \qquad \text{(A33a, b)}$$

and quadratic form (A1) in the limit $e = 0$ reduces to the following form

$$ds^2 = b^2 \left[ -dh^2 + d c^2 + A^2 \left( dq^2 + \sin^2 q \, d j^2 \right) \right]. \qquad \text{(A34)}$$

Here

$$A = \frac{e^y - e^{-y}}{2\sqrt{-k}} , \qquad y = \sqrt{-k} \; c , \qquad \text{(A35a, b)}$$

$c$ – dimensionless radial, $q$ и $j$ – angular coordinates in $R^3$.

## REFERENCES

Bettini, A. 2001, Uspekhi Fizicheskikh Nauk (Russian), **171**, 977

Branch, D. & Tammann, G. A. 1992, A.R.A. & A., **30**, 359

Coldwell, R. R. & Steinhardt, P. J. 1998, Phys. Rev., **D 57** , 6057

Friedmann, A. 1922, Zs. F. Phys., **10**, 377

Friedmann, A. 1924, Zs. F. Phys., **21**, 326

Fukuda, Y. et al. 1998, Phys. Rev. Lett., **81**, 1562

Garnavich, P. M. et al. 1998, Ap. J., **509**, 74

Koshiba, M. 2004, Uspekhi Fizicheskhih Nauk (Russian), **174**, 419

Olive, Keith A. 2003, arXiv:astro-ph/0301505

Oort, J. H. 1983, A.R.A. & A., **21**, 373

Peebles, P. G. E. 1999, Nature, **398**, 25

Perlmutter, S. et al. 1999, A. J., **517**, 565

Riess, A. G. et al. 1998, A. J., **116**, 1009

Robertson, H. P. 1933, Rev. Mod. Phys., **5**, 62

Sazhin, M. V. 2004, Uspekhi Fizicheskikh Nauk (Russian), **174**, 198

Synge, J. L. 1960, Relativity: The General Theory, (Amsterdam: North–Holland Publishing Company).

Tarakanov, P. A. 2005, in Conference Space Physics, XXXIII (St. Petersburg University, Russia), http://www.astronet.ru/db/msg/1202478

Wark, D.L. 2003, Nucl. Phys. Proc. Suppl., **117**, 164

Wang, L. et al. 2000, Ap. J., **530**, 17

S. Weinberg, S. 1972, Gravitation and Cosmology: Principles and Applications of the General Theory of Relativity (New Jork – London – Sydney – Toronto: John Wiley and Sons, Inc.).

Weinberg, S. 1989, Rev. Mod. Phys., **61**, 1